
%
%
%
%
\normalbaselineskip = 12 pt
\magnification = 1200
\hsize = 15 truecm \vsize = 22 truecm \hoffset = 1.0 truecm
\rightline{LANDAU-92-TMP-1}
\rightline{October 1992}
\medskip
\centerline{{\bf CONFORMAL BLOCKS OF COSET CONSTRUCTION:}}
\centerline{{\bf ZERO GHOST NUMBER}}
\medskip
\centerline{M.YU.LASHKEVICH\footnote{$^\dagger$}{Supported by
Landau Institute Foundation.}}
\centerline{Landau Institute for Theoretical Physics,}
\centerline{Kosygina 2, GSP 1, 117940 Moscow V-334, Russian Federation
\footnote{$^*$}{E-mail: ngm@cpuv1.net.kiae.su with note in Subject:
Lashkevich/Landau Inst.}
}
\medskip
\centerline{{\bf Abstract}}
\medskip
\par\noindent
It is shown that zero ghost conformal blocks of coset theory G/H are
determined uniquely by those of G and H theories. G/G theories are
considered as an example, their structure constants and correlation
functions on sphere are calculated.
\bigskip
\par\noindent
{\bf 1. Introduction}
\medskip\par\noindent
Conformal blocks of 2D conformal quantum field theory$^{1-5}$ are chiral
(holomorphic) $N$-point functions out of which the monodromy invariant
$N$-point correlation function is built. Conformal blocks are usually
solutions of a differential equation. For minimal models this equation
is produced by null vectors of Verma module of Virasoro algebra$^1$
or $W$ algebra.$^6$ Wess-Zumino-Witten (WZW) theories$^7$ possess
Knizhnik-Zamolodchikov equation produced by the relation between
energy-momentum tensor and Kac-Moody currents. Conformal blocks are
solutions of this equation with constraints imposed by null vectors
of Kac-Moody algebra.$^{7,8}$
\par
Coset models$^{9-21}$ obey some special features. Goddard, Kent and
Olive$^{9-10}$ (GKO) proposed them as `quotients' G/H of two
theories in sense of factorization of state spaces. Gawedzki and
Kupiainen$^{11-13}$ proposed Lagrangian theory of coset construction,
based on gauged WZW theories. BRST approach was used to investigate
states and conformal blocks.$^{14-17}$ The difficulty of coset
models is that they offen have not their own differential equations.
For example, the simplest coset model SU$(2)_N/$U$(1)$ ($Z_N$
parafermions) adopts in fact the differential equation from SU$(2)_N$
theory,$^{18}$ although it has currents. For more complex coset
theories the currents turn out essentially non-local.$^{19,20}$
\par
The other approach to coset construction consists in finding relation
between coset conformal blocks and those of G and H theories. These
relations can be obtainned in Lagrangian approach$^{13}$ or from
duality in general consideration.$^{21}$ Here this general way
independent from Lagrangian and any other details of G and H theories
is analized. The existance and uniqueness of coset is proved (Sec. 2).
The connection with BRST approach on zero-ghost-number level is
shown (Sec. 3). The results are applied to G/G coset models (Sec. 4).
Another application to bosonic representation of coset theories
can be found in Ref. 21.
\par
$Note\ added\ in\ February\ 1993.$ Recently I have learnt that
Eq. (2.3) was first supposed by M. R. Douglas$^{23}$ and proved
and investigated by M. B. Halpern and N. Obers$^{24}$. I am
very grateful to M. B. Halpern and N. Obers for calling my
attention to this.
\vskip 30 pt
\par\noindent
{\bf 2. The Main Equation}
\medskip\par\noindent
Conformal block $\Phi_{\lambda_1\mu_1,\lambda_2\mu_2}^{\lambda_3\mu_3
\lambda_4\mu_4}(\lambda_5|_{\alpha_1}^{\alpha_2};z)$ of H theory is a
holomorphic `correlation function' of four\footnote{$^a$}{Four-point conformal
blocks are considered for simplicity - the designations are overloaded
as they are. In a sense that is sufficient.} chiral fields
$\phi_{\lambda_1\mu_1}(0)$, $\phi_{\lambda_2\mu_2}(z)$,
$\phi_{\lambda_3\mu_3}(1)$, $\phi_{\lambda_4\mu_4}(\infty)$
with intermediate state from module $\lambda_5$ (Fig. 1). Here
$\lambda_i$ labels an irreducible representation ${\bf H}^H_
{\lambda_i}$ of $\hat{h}$ chiral algebra (Virasoro, Kac-Moody, $W$
etc.); $\mu_i$ labels a state in the irreducible module $\lambda_i$;
$\alpha_1=1,\cdot\cdot\cdot,N^H_{\lambda_1\lambda_2\lambda_5^+}$;
$\alpha_2=1,\cdot\cdot\cdot,N^H_{\lambda_5\lambda_3\lambda_4}$,
where $\lambda^+$ is conjugate representation to $\lambda$, and
$N^H_{\lambda\lambda'\lambda''}$ are multiplicities of fusion rules$^5$
$$
{\bf H}^H_\lambda\otimes{\bf H}^H_{\lambda'}=\bigoplus_{\lambda''}
\bigoplus_{\alpha=1}^{N^H_{\lambda\lambda'(\lambda'')^+}}
{\bf H}^H_{\lambda''}.\eqno(2.1)
$$
For SU$(2)_k$ model$^8$ and conformal models$^{1-4}$ with central charge of
Virasoro algebra $c<1$ multiplicities are $0$ (fusion is forbidden) or $1$
(fusion is permitted).
\par
For G theory any representation ${\bf H}^G_l$ of $\hat{g}$ chiral algebra
can be decomposed into the direct sum of H modules
$$
{\bf H}^G_l=\bigoplus_\lambda\bigoplus_{n=1}^{N_\lambda}
{\bf H}^H_\lambda,\eqno(2.2)
$$
where $N_\lambda$ is the multiplicity (possibly infinite)
of module ${\bf H}^H_\lambda$ in ${\bf H}^G_l$. Thus, one can designate
conformal block of G model as $F_{l_1\lambda_1n_1\mu_1,l_2\lambda_2n_2
\mu_2}^{l_3\lambda_3n_3\mu_3,l_4\lambda_4n_4\mu_4}(l_5|_{a_1}^{a_2};z)$,
where $a_1=1,\cdot\cdot\cdot,N^G_{l_1l_2l_5^+}$; $a_2=1,\cdot\cdot\cdot
,N^G_{l_5l_3l_4}$; $n_i=1,\cdot\cdot\cdot,N_{\lambda_i}$; $\mu_i$ is
again a label of a state from the H module ${\bf H}^H_{\lambda_i}$.
\par
A conformal block of coset model will be written as $\Psi_{l_1\lambda_1n_1,
l_2\lambda_2n_2}^{l_3\lambda_3n_3,l_4\lambda_4n_4}(l_5\lambda_5|
_{a_1\alpha_1}^{a_2\alpha_2};z)$.
\par
Consider the direct product of space of conformal blocks of H theory
and of that space of the coset theory. We will look for conformal blocks
of G model in this product.$^{21}$ In other words we will search them in
form
\bigskip\leftline{
$F_{l_1\lambda_1n_1\mu_1,l_2\lambda_2n_2
\mu_2}^{l_3\lambda_3n_3\mu_3,l_4\lambda_4n_4\mu_4}(l_5|_{a_1}^{a_2};z)
$}
$$
=\sum_{\lambda_5\alpha}X_{\lambda_1\lambda_2}^{\lambda_3
\lambda_4}\big(\lambda_5|_{\alpha_1\alpha'_1}^{\alpha_2\alpha'_2}\big)
\Phi_{\lambda_1\mu_1,\lambda_2\mu_2}^{\lambda_3\mu_3,\lambda_4\mu_4}
(\lambda_5|_{\alpha_1}^{\alpha_2};z)
\Psi_{l_1\lambda_1n_1,
l_2\lambda_2n_2}^{l_3\lambda_3n_3,l_4\lambda_4n_4}(l_5\lambda_5|
_{a_1\alpha'_1}^{a_2\alpha'_2};z).\eqno(2.3)
$$
Here $X_{\lambda_1\lambda_2}^{\lambda_3\lambda_4}\big(\lambda_5|
_{\alpha_1\alpha'_1}^{\alpha_2\alpha'_2}\big)$ are coefficients to be found.
These coefficients must ensure duality of the r.h.s. or, equivalently,
its right monodromy properties. Monodromy matrices are known only to
depend on fractional parts of conformal dimensions and fusion rules.$^{2,3,21}$
Thus, monodromy matrix of $\Psi_{l_1\lambda_1n_1,
l_2\lambda_2n_2}^{l_3\lambda_3n_3,l_4\lambda_4n_4}(l_5\lambda_5|
_{a_1\alpha_1}^{a_2\alpha_2};z)$ is the same  as that of the product
$$
\Phi_{\lambda_1\mu_1,\lambda_2\mu_2}^{\lambda_3\mu_3,\lambda_4\mu_4}
(\lambda_5|_{\alpha_1}^{\alpha_2};z^*)
F_{l_1\lambda_1n_1\mu_1,
l_2\lambda_2n_2\mu_2}^{l_3\lambda_3n_3\mu_3,l_4\lambda_4n_4\mu_4}(l_5|
_{a_1}^{a_2};z),\eqno(2.4)
$$
where $z^*$ is the complex conjugate to $z$. Substituting it in Eq.(2.3)
(understood now only as an equality between monodromy matrices
rather than between conformal blocks themselves)
and concelling $F$-blocks, we find that
$$
\sum_{\lambda_5\alpha}X_{\lambda_1\lambda_2}^{\lambda_3\lambda_4}
\big(\lambda_5|_{\alpha_1\alpha'_1}^{\alpha_2\alpha'_2}\big)
\Phi_{\lambda_1\mu_1,\lambda_2\mu_2}^{\lambda_3\mu_3,\lambda_4\mu_4}
(\lambda_5|_{\alpha_1}^{\alpha_2};z)
\Phi_{\lambda_1\mu_1,\lambda_2\mu_2}^{\lambda_3\mu_3,\lambda_4\mu_4}
(\lambda_5|_{\alpha'_1}^{\alpha'_2};z^*)\eqno(2.5)
$$
is monodromy invariant. But it means that the expression (2.5) can be
considered as a correlation function, and coefficients
$X_{\lambda_1\lambda_2}^{\lambda_3\lambda_4}\big(\lambda_5|
_{\alpha_1\alpha'_1}^{\alpha_2\alpha'_2}\big)$ are found by
solving Dotsenko-Fateev equation.$^{3,21}$
\par
Now we will show that Eq. (2.3) with given $X$-coefficients permits
to find conformal blocks uniquely. We will use the following matrix
designations. $F$ will be considered as a matrix
$\big( F_{l_1\lambda_1n_1,l_2\lambda_2n_2}
^{l_3\lambda_3n_3,l_4\lambda_4n_4}(z)\big)
_{ij}$ with $i=(\mu_1\mu_2\mu_3\mu_4)$, $j=(l_5a_1a_2)$; $X$ as a matrix
$\big(X_{\lambda_1\lambda_2}^{\lambda_3\lambda_4}\big)_{ij}$,
$i=(\lambda_5\alpha_1\alpha_2)$, $j=(\lambda'_5\alpha'_1\alpha'_2)$,
diagonal over $\lambda_5\lambda'_5$; $\Phi$ as
$\big(\Phi_{\lambda_1\lambda_2}^{\lambda_3\lambda_4}(z)\big)_{ij}$,
$i=(\mu_1\mu_2\mu_3\mu_4)$, $j=(\lambda_5\alpha_1\alpha_2)$;
$\Psi$ as $\big(\Psi_{l_1\lambda_1n_1,
l_2\lambda_2n_2}^{l_3\lambda_3n_3,l_4\lambda_4n_4}(z)\big)_{ij}$,
$i=(\lambda_5\alpha_1\alpha_2)$, $j=(l_5a_1a_2)$. Eq. (2.3) can be
represented in matrix form as follows
$$
F_{l_1\lambda_1n_1,l_2\lambda_2n_2}^{l_3\lambda_3n_3,l_4\lambda_4n_4}(z)=
\Phi_{\lambda_1\lambda_2}^{\lambda_3\lambda_4}(z)
X_{\lambda_1\lambda_2}^{\lambda_3\lambda_4}
\Psi_{l_1\lambda_1n_1,l_2\lambda_2n_2}^{l_3\lambda_3n_3,l_4\lambda_4n_4}(z).
\eqno(2.3a)
$$
If it will not lead to confusion we will ommit non-matrix indices
too:
$$
F(z)=\Phi(z)X\Psi(z).\eqno(2.3b).
$$
\par
Recall that conformal blocks $\Phi(z)$ are solutions of linear differential
equation with some linear constraints. Index $j=(\lambda_5\alpha_1\alpha_2)$
numerates all linearly independent solutions. Consider minimal set of
rows of $\Phi$-matrix forming some matrix $\hat{\Phi}(z)$, such that
$\Phi(z)$ is expressed algebraically through this minimal set:
$$
\Phi_{\lambda_1\lambda_2}^{\lambda_3\lambda_4}(z)=
A_{\lambda_1\lambda_2}^{\lambda_3\lambda_4}(z)
\hat{\Phi}_{\lambda_1\lambda_2}^{\lambda_3\lambda_4}(z),\eqno(2.6)
$$
where $A_{\lambda_1\lambda_2}^{\lambda_3\lambda_4}(z)$ is some matrix
function. The matrix
$\hat{\Phi}_{\lambda_1\lambda_2}^{\lambda_3\lambda_4}(z)$ is evidently
square and $A_{\lambda_1\lambda_2}^{\lambda_3\lambda_4}(z)$ is of
maximal rank.
\par
Notice that
$$
F_{l_1\lambda_1n_1,l_2\lambda_2n_2}^{l_3\lambda_3n_3,l_4\lambda_4n_4}(z)=
A_{\lambda_1\lambda_2}^{\lambda_3\lambda_4}(z)
\hat{F}_{l_1\lambda_1n_1,l_2\lambda_2n_2}^{l_3\lambda_3n_3,l_4\lambda_4n_4}(z),
\eqno(2.7)
$$
where $\hat{F}(z)$ is the matrix consisting of rows with the same
indices $i=(\mu_1\mu_2\mu_3\mu_4)$ as in $\hat{\Phi}(z)$. Indeed,
all currents act on $\Phi(z)$ as algebraical or differential operators.
Taking into account the differential equation we can present them as
purely algebraical operators forming $A(z)$ in Eq. (2.6). These very
operators are used to decompose ${\bf H}_l^G$ in ${\bf H}_\lambda^H$
in Eq. (2.2) and act on $F(z)$ (Eq. (2.7)). This claim is induced
in the accurate definition of a coset model. Particularly, if $\hat{g}$
and $\hat{h}$ are Kac-Moody algebras, induced by WZW currents,
this claim is satisfied automatically. Indeed, any current, $J(z)$, acts on
a correlation function of primary fields as
$$
\langle J(z)\phi_1(z_1)...\phi_N(z_N)\rangle
=\sum_{i=1}^N{t_i\over z-z_i}
\langle\phi_1(z_1)...\phi_N(z_N)\rangle,
$$
where $t_i$ is the representation of primary field $\phi_i(z_i)$.
If there are more than one currents, the formula is a bit more complex
because of central terms, but in any case these Ward equations have just
the same form both in H subalgebra of G WZW theory and in H WZW theory.
\par
 From Eqs. (2.3), (2.6) and (2.7) we find
$$
A(z)\hat{F}(z)=A(z)\hat{\Phi}(z)X\Psi(z),
$$
or, taking into account maximal rank of $A(z)$ in generic point,
$$
\hat{F}(z)=\hat{\Phi}(z)X\Psi(z).\eqno(2.8)
$$
We admit that $X$ is indegenerate, and it remains to prove the same
about $\Phi(z)$. But $\det\hat{\Phi}(z)$ is simply Wronskian of linearly
independent solutions. Therefore
$$
\det\hat{\Phi}(z)\neq 0\eqno(2.9)
$$
for all $z\neq 0,1,\infty$, and solution of Eq. (2.8) is unique:
$$
\Psi(z)=X^{-1}\hat{\Phi}^{-1}(z)\hat{F}(z).\eqno(2.10)
$$
This function has not any additional singular points exept $0$, $1$ and
$\infty$. Moreover, the differential equation with constraints can be
reduced to form
$$
\partial^ny(z)-\big({\alpha\over z}+{\beta\over 1-z}\big)\partial^{n-1}y(z)
+\cdot\cdot\cdot=0,
$$
where $y(z)$ is one of the components of $\Phi(z)$ and dots mean lower
derivatives. From elementary theory of differential equations we find
$$
\det\hat{\Phi}(z)=Cz^\alpha(1-z)^\beta,\eqno(2.11)
$$
where $C$ is a constant. We see that if $F(z)$ and $\Phi(z)$ are functions
of hypergeometric type,$^{2,3}$ $\Psi(z)$ is also of hypergeometric type.
\vskip 30 pt
\noindent
{\bf 3. Conjugate Theory}
\medskip\noindent
Consider the matrix
$$
\hat{M}_{\lambda_1\lambda_2}^{\lambda_3\lambda_4}(z)=
\big(X_{\lambda_1\lambda_2}^{\lambda_3\lambda_4}\big)^{-1}
\big(\hat{\Phi}_{\lambda_1\lambda_2}^{\lambda_3\lambda_4}(z)\big)^{-1}.
\eqno(3.1)
$$
Then
$$
\Psi(z)=\hat{M}(z)\hat{F}(z).\eqno(3.2)
$$
The matrix $\hat{M}(z)$ can be supplemented by zero rows to a matrix
$M(z)$ so as
$$
\Psi(z)=M(z)F(z)\eqno(3.3)
$$
(without hats). The matrix $M$ depends on choice of the minimal set
of rows of $\Phi(z)$, but $\Psi(z)$ does not. The choice of minimal
set is a kind of gauge fixing.
\par
Consider the case of Kac-Moody algebra $\hat{h}$. Knizhnik-Zamolodchikov
equation (with constraints) has the form
$$
\big(\partial-{1\over k+c_V}\theta(z)\big)\hat{\Phi}(z)=0,\eqno(3.4)
$$
where $k$ is the central charge\footnote{$^b$}{For simplicity we consider
a simple Kac-Moody algebra. Generalization to semisimple algebras is evident.}
of algebra $\hat{h}$, $c_V$ is defined by structure constants
$f^{\alpha\beta}_\gamma$ and basic inner metric $g^{\alpha\beta}$ as
$$
f^{\alpha\gamma}_\delta f^{\beta\delta}_\gamma=-c_Vg^{\alpha\beta};
$$
$\theta(z)$ is a matrix which only depends on representations $\lambda_i$
and minimal set. Then
$$
0=\partial(\hat{\Phi}X\hat{M})=\partial\hat{\Phi}\cdot X\hat{M}
+\hat{\Phi}X\cdot\partial\hat{M}
={1\over k+c_V}\theta+\hat{\Phi}X\partial\hat{M}
=\hat{\Phi}X\big(\partial\hat{M}+{1\over k+c_V}\hat{M}\theta\big),
$$
and
$$
\partial\hat{M}+{1\over k+c_V}\hat{M}\theta(z)=0,
$$
or
$$
\big(\partial+{1\over k+c_V}\theta(z)\big)\hat{M}^+(z)=0.\eqno(3.5)
$$
The sign $^+$ designates here conjugation of all representations (see
above) with transposition of matrix. Eq. (3.5) differs from Eq. (3.4)
by the sign before ${1\over k+c_V}\theta(z)$. Therefore, Eq. (3.5)
is the Knizhnik-Zamolodchikov equation for Kac-Moody algebra $\tilde{h}$
with central charge $\tilde{k}$ such that
$$
{1\over\tilde{k}+c_V}=-{1\over k+c_V},
$$
or
$$
\tilde{k}=-2c_V-k.\eqno(3.6)
$$
On the other hand, one can consider an arbitrary solution
$\tilde{\Phi}_{\lambda_1\mu_1,\lambda_2\mu_2}^{\lambda_3\mu_3,\lambda_4\mu_4}
(\lambda_5|_{\alpha_1}^{\alpha_2};z)$ of Knizhnik-Zamolodchikov equation for
Kac-Moody algebra $\tilde{h}$ with central charge $\tilde{k}$ (conjugate
theory$^{12,14-16}$). If $k>0$, then $\tilde{k}<0$ and the conjugate
theory is not rational. Thus, there is no constraints produced by null
vectors in the congugate theory. Diagonalizing $\tilde{\Phi}$ matrix
on primary fields, it is easy to prove that the solutions, which are not
linear combinations of columns of $M^+$, do not contribute
in expression $\tilde{\Phi}^+F$ (matrices are restricted on primary
fileds), and
$$
MF=\tilde{\Phi}^+F.
$$
Finally,
\bigskip\leftline{$
\Psi_{l_1\lambda_1n_1,
l_2\lambda_2n_2}^{l_3\lambda_3n_3,l_4\lambda_4n_4}(l_5\lambda_5|
_{a_1\alpha_1}^{a_2\alpha_2};z)
$}
$$
=\sum_{\mu_1\cdot\cdot\cdot\mu_4}^{prim}
F_{l_1\lambda_1n_1\mu_1,l_2\lambda_2n_2
\mu_2}^{l_3\lambda_3n_3\mu_3,l_4\lambda_4n_4\mu_4}(l_5|_{a_1}^{a_2};z)
\tilde\Phi_{\lambda^+_1\mu^+_1,\lambda^+_2\mu^+_2}
^{\lambda^+_3\mu^+_3,\lambda^+_4\mu^+_4}
(\lambda_5|_{\alpha_1}^{\alpha_2};z),\eqno(3.7)
$$
where the summation is done over components of primary with respect
to H fields. Eq. (3.7) holds evidently in general case if conjugate
theory is irrational, and not just in the case of Kac-Moody algebra $\hat{h}$.
\par
It should by noted that $\lambda_5$ in Eq. (3.5) must only sweep over
modules that are present in fusion rules of $\hat{h}$ theory. For
other $\lambda_5$ which are present in fusion rules of $\tilde{h}$ model,
the r.h.s. of Eq. (3.7) vanishes.
\par
The r.h.s. is annihilated by any current $J_n^\alpha+\tilde{J}_n^\alpha$,
$n\geq 0$ ($J_n^\alpha$ means a current of H theory and $\tilde{J}_n^\alpha$
of the conjugate theory $\tilde{H}$),
acting at any point $0$, $z$, $1$, and $\infty$. This can be expressed in
BRST form$^{14-17}$ by using ghosts $b^\alpha(z)=\sum_nb_n^\alpha z^{-n-
\Delta(b^\alpha)}$ and $c_\alpha(z)=\sum_nc_{\alpha n}z^{-n-\Delta(c_\alpha)}$
with conformal dimensions related with those of currents, $\Delta(J^\alpha)$,
as $\Delta(b^\alpha)=\Delta(J^\alpha)$ and
$\Delta(c_\alpha)=1-\Delta(J^\alpha)$. Their non-zero anticommutation
relations are given by
$$
b^\alpha(z')c_\beta(z)={\delta^\alpha_\beta\over z'-z}+O(1),\eqno(3.8a)
$$
$$
\{b^\alpha_m,c_{\beta n}\}=\delta^\alpha_\beta\delta_{m+n,0}.\eqno(3.8b)
$$
Vacuum,\footnote{$^c$}{In Ref. 17 an\-other ghost va\-cu\-um (ty\-pe II
ac\-cord\-ing to
Hu\--Yu clas\-si\-fi\-ca\-tion$^{16}$).} $|{\hbox{ }}\rangle^{\hbox{gh}}$,
satisfies the equations
$$
b_n^\alpha|{\hbox{ }}\rangle^{\hbox{gh}}=0,{\hbox{ }}n\geq 0;
$$
$$
c_{\alpha n}|{\hbox{ }}\rangle^{\hbox{gh}}=0,{\hbox{ }}n>0.\eqno(3.9)
$$
Ghost energy-momentum tensor is given by
$$
T^{\hbox{gh}}(z)=\sum_\alpha\big((\Delta(J^\alpha)-1):b^\alpha\partial
c_\alpha:
-\Delta(J^\alpha):c_\alpha\partial b^\alpha:\big).\eqno(3.10)
$$
The normal ordering is consisted with vacuum (3.9).
\par
Current algebra $\hat{h}$ has the form
\bigskip\leftline{$
J^\alpha(z')J^\beta(z)=f^{\alpha\beta}_\gamma(z',z)J^\gamma(z)
$}
$$
+\{{\hbox{central term}}\}+\{{\hbox{regular terms}}\},\eqno(3.11a)
$$
or
$$
[J^\alpha_m,J^\beta_n]=f^{\alpha\beta,m+n}_{mn,\gamma}J^\gamma_{m+n}
+\{{\hbox{central term}}\}.\eqno(3.11b)
$$
Particularly, in the case of Kac-Moody algebra
$$
f^{\alpha\beta}_\gamma(z',z)={f^{\alpha\beta}_\gamma\over z'-z},{\hbox{ }}
f^{\alpha\beta,m+n}_{mn,\gamma}=f^{\alpha\beta}_\gamma.\eqno(3.12)
$$
\par
The BRST charge is given by
$$
Q=\oint{dz\over 2\pi i}\big(c_\alpha(z)(J^\alpha(z)+\tilde{J}^\alpha(z))
-{1\over 2}\oint{dz'\over 2\pi i}f^{\alpha\beta}_\gamma(z',z)
:c_\alpha(z')c_\beta(z)b^\gamma(z):\big)
$$
$$
=\sum_mc_{\alpha ,-m}(J^\alpha_m+\tilde{J}^\alpha_m)
-{1\over 2}\sum_{m,n}f^{\alpha\beta,m+n}_{mn,\gamma}
:c_{\alpha,-m}c_{\beta,-n}b^\gamma_{m+n}:,\eqno(3.13)
$$
and the ghost number operator
$$
U=\oint{dz\over 2\pi i}:c_\alpha(z)b^\alpha(z):
=\sum_m:c_{\alpha,-m}b^\alpha_m:.\eqno(3.14)
$$
They form the usual BRST algebra
$$
\{Q,Q\}=0,{\hbox{ }}[Q,U]=-Q,{\hbox{ }}[U,U]=0.\eqno(3.15)
$$
The condition
$$
(J^\alpha_n+\tilde{J}^\alpha_n)|\phi\rangle=0,{\hbox{ }}n\geq 0\eqno(3.16)
$$
is equivalent to
$$
Q|\phi\rangle=0,{\hbox{ }}b_n^\alpha|\phi\rangle=
c_{n+1}^\alpha|\phi\rangle=0,{\hbox{ }}n\geq 0.\eqno(3.17)
$$
Hence
$$
Q|\phi\rangle=0,{\hbox{ }}U|\phi\rangle=0.\eqno(3.18)
$$
This condition is, in the general case, weaker than Eq. (3.17),
and the coset state space is a subspace of the BRST cohomology
with zero ghost number
\bigskip\leftline{$
{\bf H}^{\hbox{coset}}
=({\hbox{Ker }}Q/{\hbox{Im }}Q)\cap\bigcap_{n\geq 0}
({\hbox{Ker }}b_n^\alpha\cap{\hbox{Ker }}c_{n+1}^\alpha)
$}
$$
\subseteq H^0({\bf H}^G\otimes{\bf H}^{\tilde{H}}\otimes{\bf H}^{\hbox{gh}},Q)
=({\hbox{Ker }}Q/{\hbox{Im }}Q)\cap{\hbox{Ker }}U.\eqno(3.19)
$$
Here ${\bf H}^G$ is the state space of G theory, ${\bf H}^{\tilde{H}}$
of the theory conjugate to H, ${\bf H}^{\hbox{gh}}$ of ghosts.
\medskip\noindent
{\bf 4. A Simple Example: G/G Theory}
\medskip\noindent
In the case of G/G model the whole G module $\lambda$ turns into a single
field of coset construction. Eq. (2.3) takes the form
\bigskip
\leftline{$
F_{\lambda_1\mu_1,\lambda_2\mu_2}^{\lambda_3\mu_3,\lambda_4\mu_4}
(\lambda_5|_{\alpha_1}^{\alpha_2};z)
$}
$$
=\sum_{\beta_1\beta_2\gamma_1\gamma_2}
X_{\lambda_1\lambda_2}^{\lambda_3\lambda_4}\big(\lambda_5|
_{\beta_1\gamma_1}^{\beta_2\gamma_2}\big)
F_{\lambda_1\mu_1,\lambda_2\mu_2}^{\lambda_3\mu_3,\lambda_4\mu_4}
(\lambda_5|_{\beta_1}^{\beta_2};z)
\Psi_{\lambda_1\lambda_2}^{\lambda_3\lambda_4}(\lambda_5|
_{\alpha_1\gamma_1}^{\alpha_2\gamma_2};z),
\eqno(4.1)
$$
or
$$
F=FX\Psi.\eqno(4.1a)
$$
Hence, conformal block of the coset construction
$$
\Psi(z)=X^{-1}\eqno(4.2)
$$
does not depend on $z$. It means that the theory is topological. With the
monodromy properties of coset conformal blocks (see (2.4))
we find correlation functions
\footnote{$^d$}{The four-point correlation function can be built from
conformal blocks not only by the same $X$-constants, but by any solution
of Dotsenko-Fateev equation.$^{22}$ Nevertheless, we will analize only
the simplest case with the same $X$-constants in both coset factorizing
and pairing chiral fields.}
\bigskip\leftline
{$
\langle\lambda_1\lambda_2\lambda_3\lambda_4\rangle
$}
$$
=\sum_{\lambda_5\alpha\beta}
X_{\lambda_1\lambda_2}^{\lambda_3\lambda_4}\big(\lambda_5|
_{\alpha_1\alpha'_1}^{\alpha_2\alpha'_2}\big)
\big(X_{\lambda_1\lambda_2}^{\lambda_3\lambda_4}\big(\lambda_5|
_{\beta_1\beta'_1}^{\beta_2\beta'_2}\big)\big)^*
\Psi_{\lambda_1\lambda_2}^{\lambda_3\lambda_4}(\lambda_5)
_{\beta_1\alpha_1}^{\beta_2\alpha_2}
\big(\Psi_{\lambda_1\lambda_2}^{\lambda_3\lambda_4}(\lambda_5)
_{\beta'_1\alpha'_1}^{\beta'_2\alpha'_2}\big)^*
$$
$$
={\hbox{Tr }}(\Psi X\Psi^\dagger X^\dagger)={\hbox{Tr }}1.
$$
Finally,
$$
\langle\lambda_1\lambda_2\lambda_3\lambda_4\rangle
=N_{\lambda_1\lambda_2}^{\lambda_3\lambda_4}\equiv
\sum_{\lambda_5}N_{\lambda_1\lambda_2\lambda_5^+}
N_{\lambda_5\lambda_3\lambda_4},\eqno(4.3)
$$
where $N_{\lambda_1\lambda_2}^{\lambda_3\lambda_4}$ is the
number of `intermediate states' $\lambda_5$ (including multiplicities)
according to fusion rules of initial G model. These correlation
function are dual (field algebra is associative). Indeed, the number
in the r.h.s. of Eq. (4.3) is simply the number linearly independent
solutions of the differential equation with constraints.$^1$
It does not, of course, depend on choice of basic functions
(see also Ref. 5):
$$
\sum_{\lambda_5}N_{\lambda_1\lambda_2\lambda_5^+}
N_{\lambda_5\lambda_3\lambda_4}
=\sum_{\lambda_6}N_{\lambda_1\lambda_3\lambda_6^+}
N_{\lambda_6\lambda_2\lambda_4}.\eqno(4.4)
$$
Structure constants are given by
$$
\langle\lambda_1\lambda_2\lambda_3\rangle=N_{\lambda_1\lambda_2\lambda_3}.
\eqno(4.5)
$$
At last, $n$-point correlation functions on sphere are given by (Fig. 2)
$$
\langle\lambda_1\cdot\cdot\cdot\lambda_n\rangle
=\sum_{\Lambda_1\cdot\cdot\cdot\Lambda_{n-3}}
N_{\lambda_1\lambda_2\Lambda_1^+}N_{\Lambda_1\lambda_3\Lambda_2^+}
\cdot\cdot\cdot N_{\Lambda_{n-3}\lambda_{n-1}\lambda_n}.
\eqno(4.6)
$$
\medskip\noindent
{\bf 5. Conclusion}
\medskip\noindent
It has been shown that conformal blocks of coset model G/H are uniquely
determined by those of G and H models. This construction of conformal
blocks is equivalent to BRST construction in the ghostless sector.
\par
As an example of application of this approach G/G coset models are
considered. These theories are topological ones and their correlation
functions turn out simply numbers of intermediate states.
\par
To conclude, we present two conjectures concerning non-zero ghost numbers.
\par
{\bf Conjecture 1.} All nonvanishing correlation functions of states with
non-zero ghost numbers coincide with those of corresponding zero-gost states.
\par
{\bf Conjecture 2.} For SU$(2)_k/$SU$(2)_k$ coset for rational $k$
and Virasoro/Vi\-ra\-so\-ro for minimal models states are labelled by
pairs $(\lambda,h)$, where $h$ is ghost number. Structure constants
are given by
$$
\langle(\lambda_1,h_1)(\lambda_2,h_2)(\lambda_3,h_3)\rangle
=N_{\lambda_1\lambda_2\lambda_3}\delta_{h_1+h_2+h_3,0},
$$
and correlation functions are
\bigskip\leftline
{$
\langle(\lambda_1,h_1)\cdot\cdot\cdot(\lambda_n,h_n)\rangle
$}
$$
=\sum_{\Lambda_1\cdot\cdot\cdot\Lambda_{n-3}}
\langle(\lambda_1,h_1)(\lambda_2,h_2)(\Lambda_1,-h_1-h_2)\rangle
\langle(\Lambda_1,h_1+h_2)(\lambda_3,h_3)(\Lambda_2,-h_1-h_2-h_3)\rangle
$$
$$
\cdot...
\langle(\Lambda_{n-3},h_1+\cdot\cdot\cdot+h_{n-2})
(\lambda_{n-1},h_{n-1})(\lambda_n,h_n)\rangle.
$$
\medskip
\par
The author is grateful to Vl.S.Dotsenko for very fruitful discutions.
\bigskip
\par\noindent
{\bf References}
\par\noindent
1. A.A.Belavin, A.M.Polyakov and A.B.Zamolodchikov, $Nucl$. $Phys$.
{\bf B241} (1984) 333
\par\noindent
2. Vl.S.Dotsenko and V.A.Fateev, $Nucl$. $Phys$.
{\bf B240 [FS12]} (1984) 312
\par\noindent
3. Vl.S.Dotsenko and V.A.Fateev, $Nucl$. $Phys$. {\bf B251 [FS13]} (1985) 691
\par\noindent
4. G.Felder, $Nucl$. $Phys$. {\bf B317} (1989) 215
\par\noindent
5. G.Felder, J.Fr\"olich and G.Keller, $Commun$. $Math$. $Phys$.
{\bf 130} (1990) 1
\par\noindent
6. V.A,Fateev and A.B.Zamolodchikov, $Nucl$. $Phys$. {\bf B280 [FS18]} (1987)
644
\par\noindent
7. V.G.Knizhnik and A.B.Zamolodchikov, $Nucl$. $Phys$. {\bf B247} (1984) 83
\par\noindent
8. A.B.Zamolodchikov and V.A,Fateev, $Sov$. $J$. $Nucl$. $Phys$.
{\bf 43} (1986) iss. 9, 1031
\par\noindent
9. P.Goddard, A.Kent and D.Olive, $Phys$. $Lett$. {\bf 152} (1985) 88
\par\noindent
10. P.Goddard, A.Kent and D.Olive, $Commun$. $Math$. $Phys$.
{\bf 103} (1986) 105
11. K.Gawedzki and A.Kupiainen, $Phys$. $Lett$. {\bf B215} (1988) 119
\par\noindent
12. K.Gawedzki and A.Kupiainen, Institut des Hautes Etudes Scientifiques
\par\noindent
preprint IHES/P/88/45 (September 1988)
\par\noindent
13. K.Gawedzki, Institut des Hautes Etudes Scientifiques
preprint
\par\noindent IHES/P/89/53 (August 1989)
\par\noindent
14. G.Momma, $Prog$. $Theor$. $Phys$. {\bf 85} (1991) 371
\par\noindent
15. H.-L.Hu and M.Yu, preprint BIHEP-TH-92-22, AS-ITP-92-23,
CCAST-92-21 (April 1992)
\par\noindent
16. H.-L.Hu and M.Yu, preprint BIHEP-TH-92-35, AS-ITP-92-32 (May 1992)
\par\noindent
17. M.Yu.Lashkevich, to be published in $Int$. $J$. $Mod$. $Phys$. {\bf A}
\par\noindent
18. A.B.Zamolodchikov and V.A.Fateev, $Sov$. $Phys$. $JETP$,
{\bf 62} (1985) 215
\par\noindent
19. K.Bardakci, M.Crescimanno and S.H.Hotes, $Nucl$. $Phys$.
{\bf B349} (1991) 439
\par\noindent
20. K.Bardakci, preprint LBL-31686, UCB-PTH-92/02 (January 1992)
\par\noindent
21. M.Yu.Lashkevich, submitted to $Mod$. $Phys$. $Lett$. {\bf A}
\par\noindent
22. A.Cappelli, C.Itzykson, J.B.Zuber, $Nucl$. $Phys$.
{\bf B280 [FS18]} (1987) 445
\par\noindent
23. M. R. Douglas, preprint CALT-68-1453, 1987
\par\noindent
24. M. B. Halpern and N. Obers, preprint LBI-32619,
USB-PTH-92-24, BONN-HE-92/21, July 1992, hep/th-9207071
\end